\documentclass{article}
\usepackage{authblk}
\usepackage{bbding}
\usepackage{caption}
\usepackage{graphicx} 
\usepackage{geometry}
\usepackage{listings}
\usepackage[hyphens]{url}
\usepackage[section]{placeins}
\usepackage{svg}
\usepackage[english]{babel}
\usepackage[nottoc]{tocbibind}

\graphicspath{{./figs/}}

\title{An open framework for archival, reproducible, and transparent science}
\author[1]{Sabar Dasgupta}
\author[1,2,3,4]{Paul Nuyujukian}
\affil[1]{Department of Bioengineering}
\affil[2]{Department of Neurosurgery}
\affil[3]{Department of Electrical Engineering}
\affil[4]{Wu Tsai Neurosciences Institute}
\affil[ ]{Stanford University}

\begin{document}

\maketitle

\section{Abstract}

Digital computational outputs are now ubiquitous in the research workflow and the way in which these data are stored and cataloged is becoming more standardized across fields.
However, even with accessible data and code, the barrier to recreating figures and reproducing scientific findings remains high.
One element generally missing is the computing environment and associated pipelines in which the data and code are executed to generate figures.
The archival, reproducible, and transparent science (ARTS) open framework incorporates containers, version control systems, and persistent archives through which all data, code, and figures related to a research project can be stored together, easily recreated, and serve as an accessible platform for long-term sharing and validation.
If the underlying principles behind this framework are broadly adopted, it will improve the reproducibility and transparency of research.

\section{Introduction}

Over the past few decades, researchers have recognized the large effort required to replicate experiments and the rate at which these replication studies produce negative results \cite{ioannidis_why_2005}. 
Dubbed a ``replicability crisis'' \cite{pashler_is_2012}, the current paradigm for communicating experimental methods and distributing research outputs impedes future collaboration and the development of public trust in findings. 
To improve replicability, both the processes for data collection and those for creating reproducible analyses must be improved. 
Reproducibility is the primary focus of this writing with the goal of minimizing the time, effort, and expertise required to analyze data from a study beyond its published manuscript \cite{heil_reproducibility_2021}. 

In this writing and in the context of computational workflows that perform analysis with code on data to generate results, \textbf{reproducibility} is defined as the measure of ease by which one can run an equivalent workflow with original data to reproduce published results. 
Then, \textbf{replicability} is the measure of ease by which an equivalent workflow can be run with \textit{independently collected} data to verify published results \cite{19_repr_committees}. 
Research that is replicable is reproducible by default so long as the original data is available and usable, but the converse is not necessarily true.
Replicability sets a higher bar due to the cost associated with collecting data to repeat experiments.
At a minimum, research findings should be easily reproducible to foster collaboration and knowledge access, lower the barrier to building on an existing piece of work, and comply with new funding mandates.

By the end of of 2025, research funded by the US federal government must make data accessible at the time of publishing \cite{22_OSTP, 22_OSTP_news}.
Each funding agency has their own specific policies for data sharing, but all focus attention on timely release of data in a findable manner \cite{23_NIH_Policy,office_of_the_director_national_institutes_of_health_not-od-25-047_nodate,us_department_of_energy_scientific_nodate,national_aeronautics_and_space_administration_nasas_2025,department_of_defense_free_nodate,us_national_science_foundation_proposal_nodate}. 
While these policies offer guidance for researchers pursuing collaboration and open science, they do not outline \textit{how} data, and more importantly, entire workflows should be shared and made accessible. 
The details around implementing computational best practices are left up to researchers, which is generally outside their scope of expertise and takes the focus away from core scientific work. 

There are a number of standards put out by the scientific community to address these concerns ranging from generic (and not necessarily research-specific) to highly focused on one field of research. 
Data standards such as the FAIR Guiding Principles \cite{wilkinson_fair_2016} have gained popularity and are supported by open systems such as DataFed \cite{DataFed}.
While FAIR sets a high standard regarding the sharing of data, it alone does not assure reproducibility as it is only starting to be applied to computation \cite{wilkinson_applying_2025}.
Standards for reproducibility must advise on not only how data is stored, but also how it is manipulated and presented in analysis. 

\begin{figure}[ht!]
\includegraphics[scale=0.75]{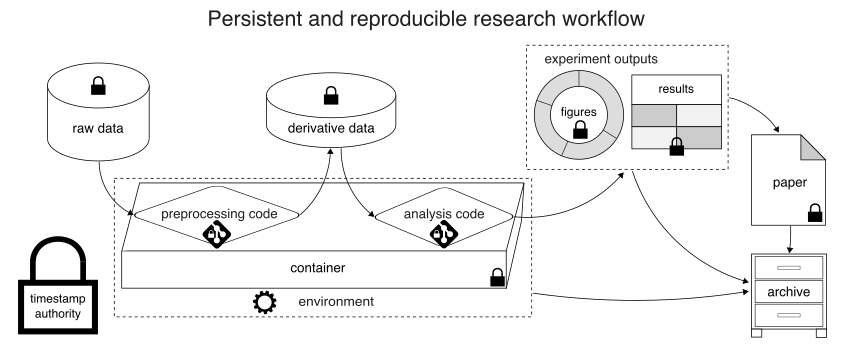}
\caption{The research workflow involves collecting raw data, transforming it into derivative data, and producing experimental outputs using code.
Ideally, all pieces of the workflow should be archived as research artifacts—the data, computation, environment, experiment outputs, and associated manuscript.
A signature of each piece of data can then be kept track of using a timestamp authority to keep a trusted historical record.}
\label{fig:workflow}
\end{figure}

Frameworks such as Spyglass (for neuroscience) or Galaxy (for bioinformatics) recommend field-specific workflows with associated software and are an excellent step towards integrating various software frameworks, but can be fragile as they rely on the permanence and compatibility of external services, software repositories, and APIs \cite{spyglass, goecks_galaxy_2010}. 
Journals have also recognized the importance software quality and availability best practices \cite{katz_publish_2018}. 
Specifically, both Nature and Science journals now prohibit the use of GitHub for code depositions due to its mutability \cite{nature_portfolio_reporting_nodate, science_journals_science_nodate} and Cell strongly recommends archives that support digital longevity \cite{cell_press_resource_nodate}.
At the time of this writing, even data-focused journals such as Nature Scientific Data provides comprehensive requirements and recommendations for deposition of data and code, but does not prescribe methods for sharing computing environments to promote accessible pipeline reproducibility \cite{scientific_data_data_nodate}.

It is becoming widely known that consistent computing environments are a  crucial piece of scientific reproducibility \cite{trisovic_large-scale_2022, ivie_reproducibility_2019, wonsil_reproducibility_2023, trisovic_advancing_2020, wang_restoring_2020, ziemann_five_2023}.
Software-as-a-Service (SaaS) tools such as Code Ocean and Brainlife offer reproducibility and analysis pipeline visualization in a standardized environment. \cite{code_ocean_computational_nodate, hayashi_brainlifeio_2024}. 
Code Ocean is hosted exclusively via a cloud provider which offers simplicity and reliability at the cost of openness.
However, Code Ocean does allow exporting of Containerfiles—the files that define the workflow environment—so that workflow build processes can be run outside of their platform. 
While Code Ocean is a useful tool for increasing reproducibility for the journal review process \cite{cheifet_promoting_2021}, it does not define an open standard and is a centralized, commercial solution that does not fully incentivize the sharing of environments throughout the research lifecycle.
Brainlife does offer the ability to run jobs on local computing hardware, but is neuroscience-specific.
The Whole Tail is yet another promising initiative and stresses the importance of cataloging software environments, but does not appear to have an active community at the time of writing   \cite{brinckman_computing_2019}.
The calkit Python package also serves as a useful reproducibility tool and acts as a wrapper around some of the technologies discussed later on \cite{calkit_calkit_nodate}.
With the potential exception of the STARS framework for discrete event simulation healthcare models \cite{monks_towards_2024}, existing environment-sharing solutions do not yet define an \textit{open framework} that aims to minimize compliance burden and the time and effort required to reproduce a research workflow many years down the line. 
Neither highly opinionated field-specific frameworks nor existing generic research workflow aids provide open, end-to-end support at the computational environment level for a wide array of fields.

Heil et al. introduce a medal scale pertaining to computational life science workflows that recommends automating dependency installation and describes a gold standard where analyses are reproducible with a signal command \cite{heil_reproducibility_2021}. 
Unless these dependencies are versioned and running in a consistent environment, there is no guarantee that they will produce deterministic outputs or even run without error in the future.
While reproduction effort is a crucial metric, a further key consideration is how far into the future the workflow is able to be run in one step. 
To truly meet this gold standard the complete software pipelines associated with experiment analyses must be captured as a \textit{container image}—a binary file that bundles all dependencies, including the operating system, needed to recreate the experiment outputs.
A comprehensive source of guidance related to these concepts is available in the Turing Way handbook \cite{the_turing_way_community_turing_2025}.

Until recently, reproducing entire research workflows was difficult, if not impossible, without locking in to specific computational platforms. 
Most published depositions do not specify a complete computing environment which leaves the work of setting up a working environment to researchers attempting to reproduce findings.
Current approaches also assume that software library repositories such as PyPI for Python or CRAN for R can serve as scientific archives and will continue to exist in their current state decades into the future.
Containers, potentially a pillar of scientific computing, offer a solution to environment management that clearly defines entire workflows while being relatively operating system agnostic across local and cloud settings \cite{nuyujukian_leveraging_2023}. 
This writing describes a framework which aims to be a flexible blueprint that is compatible with as many existing research workflows as possible.
Specific recommendations are made that encourage reproducibility and define a compatibility layer for those looking to verify results and collaborate. 
An example deposition is also provided that demonstrates use of the framework in a signal processing context using Python.

\section{Results}

\subsection{Framework Definition}

This section defines the \textbf{A}rchival, \textbf{R}eproducible, and \textbf{T}ransparent {\textbf{S}}cience (\textbf{ARTS}) open framework for accessibly sharing research involving data collection and computational analyses. 
The goal of this framework is to provide those performing computational analyses that produce research outputs (in this context, digital artifacts) a set of best practices for designing computational research efforts, archiving data and results, sharing said artifacts, and making it simple for others to reproduce and extend results within a consistent analysis environment. 
The framework can be used collaboratively, incorporating edits and sharing across multiple individuals and groups.
Pieces of the ARTS open framework may also be omitted to define best practices for partial workflows—for example, archiving just datasets, archiving code that operates on already-archived datasets, or archiving a standalone computational environment that is used in many experiments. 

For a research output that includes data and analysis code to be maximally compatible with the framework, it must include the following: 

\begin{itemize}
    \item Research outputs deposited in an accessible, persistent, and trusted archive
    \item All analysis code tracked under version control
    \item Data and code provenance timestamped (e.g., RFC 3161)
    \item Clear documentation 
    \item Configuration file and corresponding file structure
    \item Container definition and associated image
    \item Appropriate data and software licenses 

\end{itemize}

The points outlined above and described in detail below should ideally be incorporated early into the research lifecycle where possible.
That way, data collection and analysis steps can benefit from automation and archiving the research deposition poses a minimal burden.
Further, having a portable deposition that can be uploaded to multiple archives makes research outputs more resilient and having a unified file structure makes outputs more usable.

\subsection{Archives}

One goal of archiving experimental results is to make all aspects of analysis available to future experimenters for decades to come: processing data, running models, and generating figures. 
In order to achieve this, it is vital that the archive service housing the research output be freely accessible as far as possible into the future, or that depositions are able to be archived in a decentralized manner.
When choosing an ARTS-compatible archive, there a few key considerations:

Archives must be publicly \textbf{accessible}, meaning that repositories are available for download without paywalls or logins except when necessary to safeguard sensitive information such as Protected Health Information \cite{scientific_data_data_nodate}. 
Accessible archives must also support a variety of open licenses and allow works to be searchable by included metadata like keywords.

Archives must allow for \textbf{persistent} storage of artifacts (e.g., files and directories). 
Once results are published, they should remain immutable (i.e., cannot be quietly changed) and allowances for future updates or edits must result in an immutable history that includes the original publication. 
If history is not immutable, trusted timestamping information must be provided (an example is described below). 
Persistent archives must also support persistent identifiers of some kind. 
Ideally this should be a digital object identifier (DOI), but a persistent URL (PURL) that is not a DOI may suffice.
If the archive does not support DOIs, published manuscripts describing the data should be minted with a DOI and include relevant PURLs.
Centralized persistent archive services should also store a copy of data across multiple geographic regions so that they are resilient to external events that might result in data loss such as natural disasters or political events. 

Archives must also be \textbf{trusted}. 
Requirements for trust cannot be singularly defined in a broad sense, but at a minimum, an archive service should be established with long-term financial support and a dedicated community.
Archives backed by noncommercial institutions are also preferable as they are less likely to change their policies due to economic pressures.
Providing for an optional confidential review process of artifacts also promotes trust in the archive and the data it houses \cite{scientific_data_data_nodate}. 

\begin{figure}[ht!]
\includegraphics[scale=0.27]{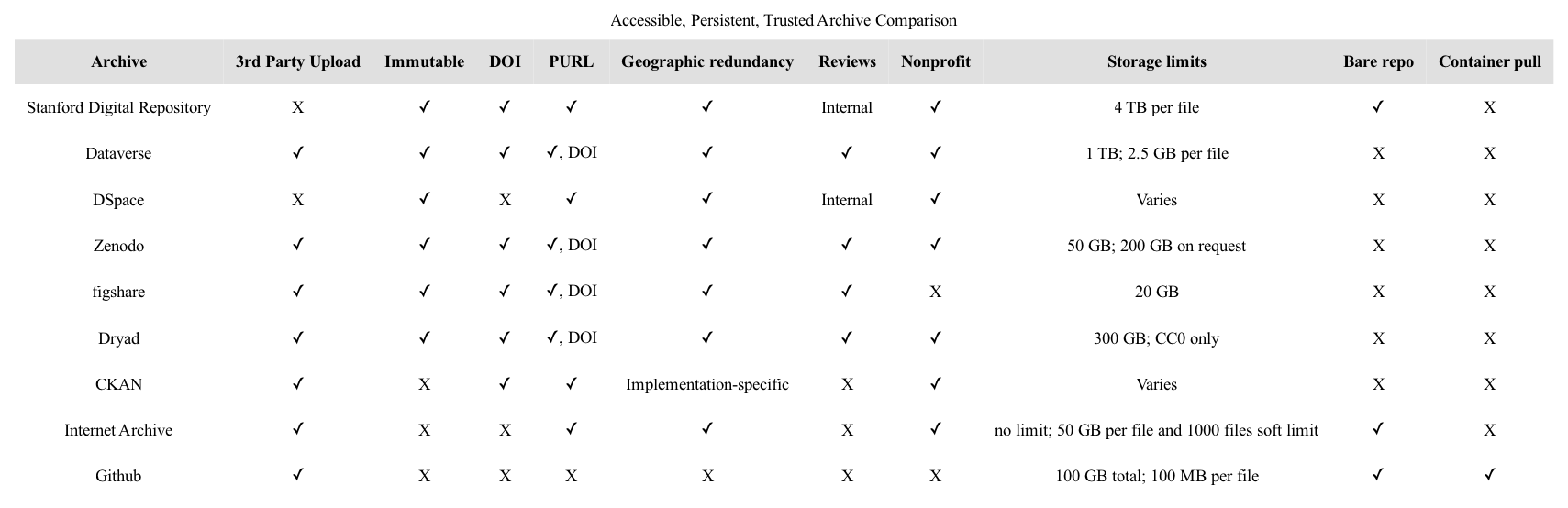}
\caption[]{No popular research archive currently supports all aspects of the ARTS open framework, but Zenodo supports all aspects outside of hosting a container registry and allowing direct \texttt{git clone}s of analysis code. Each column (besides storage limits) defines a question with checks and X's indicating yes and no, respectively. Columns definitions are as follows:

\textit{3rd party upload}: Does the archive allow anyone to upload data?

\textit{Immutable}: Once published, is deposition content locked from edits without provenance information?

\textit{DOI}: Does the archive supply a Digital Object Identifier for uploads?

\textit{PURL}: Does the archive supply a persistent URL for uploads (this may be a DOI)?

\textit{Geographic redundancy}: Is data uploaded to the archive backed up in separate geographic locations?

\textit{Reviews}: Does the archive provide a process for privately and confidentially sharing uploads to reviewers before publishing?

\textit{Nonprofit}: Does the maintainer of the archive operate under nonprofit status?

\textit{Storage limits}: Maximum file and deposition size as well as any other deposition limits.

\textit{Bare repo}: Does the archive allow depositions to be directly git cloned?

\textit{Container pull}: Does the archive implement a container registry that allows direct image pulls from container engines?
}
\label{fig:archives}
\end{figure}

Some academic institutions offer in-house archive solutions: the Dataverse Project originally developed at Harvard \cite{noauthor_dataverseorg_nodate, the_president__fellows_of_harvard_college_harvard_nodate}, DSpace developed between MIT and HP \cite{dspace_dspace_nodate}, and the Stanford Digital Repository (SDR) \cite{stanford_university_libraries_stanford_nodate} are some examples of archives that promote open access and provide persistent identifiers.
DSpace and Dataverse are open-source tools with instances hosted by many partner organizations while the SDR is only available to Stanford affiliates.
There are also popular and journal-recommended third-party solutions such as Zenodo \cite{noauthor_zenodo_nodate}, Dryad \cite{dryad_dryad_nodate}, or figshare \cite{noauthor_figshare_nodate} if a field-specific data repository cannot be found.
Dryad allows up to 300GB of storage per submission with a publication fee, while Zenodo and figshare do not require a fee to upload data, but have stricter storage limits. 

A final consideration when archiving is the ease by which content can be uploaded by researchers and then downloaded by interested parties.
The upload process should be as simple and automated as possible and ideally mirror how researchers already transfer their data between storage options. 
The Internet Archive, for example, supports uploads of data with simple file management tools like rclone \cite{internet_archive_internet_nodate}, while the Dataverse Project provides ingestion plugins for popular data management tools like iRODS, Globus, and DataLad \cite{the_dataverse_project_integrations_nodate}.
Allowing researchers to mint DOIs before upload so they may include them in the first version of their deposition is also helpful. 
In order to support accessible downloads, archives should ideally preserve artifact file structure via a static URL path and common URL prefix for the entire deposition so that files can be cloned via git as discussed in the next section. 
Stanford's Digital Repository exposes raw file structure in this way and the current best public alternative for supporting this feature is the Internet Archive, which unfortunately does not provide immutability or DOI support. 
Zenodo is overall the best choice for ARTS-compatible deposition at the moment even though it obfuscates individual file paths as a zip archive.
A gold standard archive service would also host a container registry in order to centralize and make accessible computational environments used to process deposited data.
A full visual comparison of the discussed archive options is provided in Figure \ref{fig:archives}.

\subsection{Software version control}

Git is a ``fast, scalable, distributed revision control system'' \cite{git_git_nodate} and is the most popular among version control tools which allow researchers to keep a record of experimental code and supporting files \cite{nuyujukian_leveraging_2023}. 
Like the Linux kernel, Git is licensed mostly under the open source GPLv2 license and is available without cost \cite{git_free_nodate}.
Git excels at keeping track of code and other lightweight, human readable files, but is generally not advised for use and tracking of large data files.

Experiment analysis code directories should be initialized as git repositories using the \texttt{git init} command and changes kept track over time as commits using \texttt{git add} and \texttt{git commit}. 
Researchers should be careful to not commit sensitive information such as API keys and secrets. 
Local environment files containing these values may be added to a \texttt{.gitignore} file so that git never commits them.

Git offers the \texttt{git clone} command to copy an entire git tree (files and history) into a new location. 
Accessing experiment code should be done by running this command and is often done by supplying a link to a git-hosting service such as GitHub. 
This for-profit service, now owned by Microsoft, does not provide persistent URLs or immutable repositories, but has made a commitment to archiving open source repositories \cite{github_about_nodate}, has taken specific actions to this effect \cite{github_archive_program_preserving_nodate}, and provides integrations for archiving repositories to Zenodo and figshare \cite{github_referencing_nodate}. 
That said, repository owners on GitHub may delete the repository or edit its history, breaking published links and calling provenance into question. 

Ideally, code should be directly downloadable from the archive using \texttt{git clone}. 
This is possible by placing a bare git repository (the \texttt{.git} folder created with \texttt{git init --bare}) at the top-level of a desired repository to share.
Most repository web servers will not support the Git HTTP smart service protocol extensions \cite{git_protocols_nodate}, and thus repositories should also be prepared with the \texttt{git update-server-info} command before deposition \cite{git_git-update-server-info_nodate}.
This functionality is currently supported by the Internet Archive and the Stanford Digital Repository.

\subsection{Containers}

The backbone of reproducibility in the ARTS open framework is the packaging of the workflow inside a container. 
Containers offer the promise of executing the same environment on multiple host platforms and are used by services such as Binder to provide a shared computing environment \cite{jupyter_binder_2018}. 
Their use is growing across scientific fields especially to drive collaborative and reproducible efforts \cite{moreau_containers_2023}.

Containers are executed in and managed by container engines. 
No specific container engine is prescribed by ARTS, but engines supporting Open Container Initiative (OCI)-compatible images such as Podman or Docker are recommended \cite{the_linux_foundation_open_nodate}. 
This ensures that as the technology evolves, there will be a way to run the container in the future even if the specific runtime used by the original authors is no longer supported.

Containers are specified by Containerfiles which define any relevant predefined base environment, commands to run and file structures to set up in that environment, and a jumping off point (i.e., default execution script) into the environment (entrypoint). 
Containers can take environment variables as inputs during runtime—in this case the computational configuration \texttt{config.env} file defined below. 
Files in the local file system can also be made accessible within the container by using volumes to mount a directory from the local file system within the container.

To fully conform with the ARTS open framework, the provided container should support the following workflows:

\begin{itemize}
    \item run analysis and generate static research outputs (e.g., figures or numerical results)
    \item run interactive development environment (e.g., jupyter server)
    \item run interactive command line interface (e.g., IPython kernel or IRkernel)
\end{itemize}

Ideally, these scripts should be implemented early in the research lifecycle to provide convenience during the exploratory and iterative phases of research instead of acting as a barrier to submission.
Additional commands provided with the container to perform actions like generating interactive plots and complete manuscript figures, or running data collection pipelines should also be included when possible.

Finally, container \textit{images} defining the computation environment must be published along with other research outputs. 
Container images are compiled, executable packages that are the end result of building a Containerfile. 
It is recommended to compile container images on an x86\_64 host machine and release it for that architecture at a minimum, but other host architecture images may be used and released as well (e.g., i386, arm64).

\subsection{Trusted timestamping}

Trusted timestamping (RFC3161/RFC5816) is a technique that provides secure tracking of provenance information (i.e., creation and modification dates) for any digital content including data and code \cite{nuyujukian_leveraging_2023}. 
Since data and code have different life cycles, the trusted timestamping process differs between them.

Data should be timestamped during the collection process.
A reference implementation is available using the scripts in the Trusted Timestamping Framework for Scientific Research \cite{brain_interfacing_laboratory_biltimestamping_2025} or manually with the associated web tool \cite{brain_interfacing_lab_trusted_2025}.
The framework generates a timestamps.json file which should be included at the top-level directory where timestamped data is present.
An example timestamped dataset is provided for reference \cite{nuyujukian_w241130_2024}. 

By applying trusted timestamping to a git code repository, the commit history can be validated even if the code hosting service does not offer immutable repositories. 
The process of adding trusted timestamping to experimental git repositories is simple using the framework referenced above.
The relevant \texttt{post-commit} file is added to the \texttt{.git} directory of the repository and all subsequent commits result in a timestamp record being generated and stored in a publicly available record.
This is a straightforward and open-source mechanism to create an electronic lab notebook.

\subsection{Documentation}

At the very least, basic documentation included as a \texttt{README} file should be present at the top level of the deposition directory detailing an overview of the dataset and its formats, instructions on how to run containerized code, and information on the outputs produced by this code.
Docstrings and concise, descriptive comments within code files are also highly recommended.

From the standpoint of reproducibility, it is most important that the instructions to setup and run analysis on the provided data are clear and simple. 
Recommended container dependencies (e.g., docker/podman version, tested architectures, etc.) should be provided as well as any scripts for convenience.

From the standpoint of replicability, clear methodology defining the data collection process should also be present. 
The fine details of this process, if included in research articles, are often left to supplementary sections and not usually accompanied by data collection pipelines. 
Artifact documentation should include technical details around data collection such as experimental setup, materials (including specific hardware) needed, and any related code.
To the extent possible, all code should also be well documented in accordance with the respective programming language guidelines.

\subsection{Configuration and file structure}

An environment variables file named \texttt{config.env} or \texttt{.config.env} \textit{must} be present at the top level folder of the deposition or under a folder named \texttt{arts/} to be compatible with the ARTS open framework. 
This file defines where code, data, and research outputs are located, locations for other required files, and any other parameters that may be passed to analysis or data collection code. 
All other files and folders, while recommended to include when relevant, are optional and may be omitted.

The following is an example default environment variables file specifying a recommended set of file and folder locations for an ARTS-compatible research output.
Path variables use Unix-like syntax and should specify relative paths with respect to the top-level directory: 

\begin{lstlisting}
ARTS_RAW_DATA_PATH=data/raw/
ARTS_DERIV_DATA_PATH=data/deriv/
ARTS_CODE_PATH=code/
ARTS_ENV_PATH=env/
ARTS_OUTPUT_PATH=output/
ARTS_LICENSE_PATH=LICENSE
ARTS_README_PATH=README.md
ARTS_RUN_CMD=./run.sh
ARTS_SETUP_CMD=./setup.sh 
\end{lstlisting}

Researchers are responsible for using these environment variables in their code so that changing them has an affect on analysis reruns.
In the ideal case, where this framework is incorporated early on in the research lifecycle, these path variables point to local resources and can later be updated to reflect archive deposition compatible paths.
If the downstream code uses these variables for all path definitions, then the code will successfully port to any environment or archive that the deposition may be moved to.
Omitting an environment variable or setting it to an empty string (e.g., \verb|ARTS_CODE_PATH=| \hspace{0pt}) indicates that the corresponding file or folder is not present in the deposition.
In some circumstances, researchers may also want to provide a PURL or other resolvable value in a path variable.
For example, if new features are being derived from an existing raw dataset, the \verb|ARTS_RAW_DATA_PATH| may be set to an external DOI or PURL where the data is accessible, so long as the setup and run commands support using an external URL.

In the case where a study includes multiple git repositories, each git repository can be included in a separate folder under the \texttt{code} directory.
Then, if a top-level git repository is defined for the deposition, the \texttt{code/} directory can be placed in a \texttt{.gitignore} file to avoid clashing git repositories.
Alternatively, git submodules may be used.

\begin{figure}[ht!]
\includegraphics[scale=0.5]{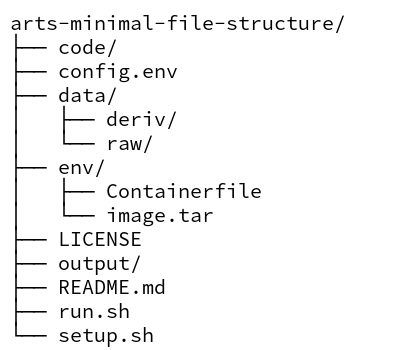}
\caption{Minimal example file structure compatible with the ARTS open framework.}
\label{fig:file_structure_min}
\end{figure}

The presence of a \texttt{config.env} greatly simplifies the reproduction process by allowing validators to rerun analyses and recreate the \texttt{output/} directory with the one-line \verb|ARTS_RUN_CMD|.
Replication is also largely simplified for results that are reproducible through the ARTS open framework. 
After running a new experiment to collect independent data, the value for \verb|ARTS_RAW_DATA_PATH| can be updated to run analyses on these data and complete the replication.
Incorporating top-level environment variables during the data collection and analysis development processes is also a recommended best practice as it provides a unified interface for parameter tuning and makes code more accessible.
For example, when archiving sensitive data such as PHI or trade secrets, access controls such as authentication are necessary, but can be accomplished via an API key supplied as an environment variable.

\subsection{Licensing}

Licenses define what rights you give others to use, modify, and share your code \textit{and data}.
By default in most jurisdictions, all rights to use novel creations are reserved and so it is necessary to spell out what uses are allowed \cite{st_laurent_understanding_2004}. 
Omitting an appropriate license from an otherwise accessible deposition will make it unusable.
Open source code and open content licenses, as opposed to proprietary or noncommercial licenses, give free access to use, modify, and share content \cite{open_source_initiative_open_2007, 19_committees_understanding}. 
Common choices for open source code licenses include permissive licenses such as BSD-3, MIT, and Apache-2.0. 
Share-alike or ``copyleft'' licenses such as the GPL family are also a popular choice as they require modified code to be distributed with a similar license and therefore encourage return contributions from collaborators \cite{engelfriet_choosing_2010}. 

As for data and other content such as documentation, Creative Commons (CC) provides three licenses that are widely used and recommended. 
Open Data Commons (ODC) also provides a set of similar open licenses that are aimed specifically at data, but may not be suitable for other content \cite{open_knowledge_foundation_conformant_nodate}.
The most recent versions of the CC licenses are recommended over ODC licenses as they provide broader coverage and permissions \cite{creative_commons_data_nodate}.
For placing data in the public domain without restriction, the CC0 license should be used, and is required for work directly generated by, or involving as authors, US federal agencies and staff. 
If attribution to the original authors is required during redistribution, the CC BY license is suitable. 
And if an attribution and share-alike clause is desired, there is the CC BY-SA license \cite{creative_commons_legal_nodate}. 
Unfortunately, none of these licenses provide terms for data provenance beyond attribution. 
For example, even a dataset released under the CC BY-SA license can be downloaded, slightly modified, and reuploaded as a purported exact copy of the original so long as it is released with attribution information and a copy of the license. 
Therefore, it is important to release attestable provenance information along with datasets as described in the trusted timestamping section. 

\begin{figure}[ht!]
\centering

\includegraphics[scale=0.7]{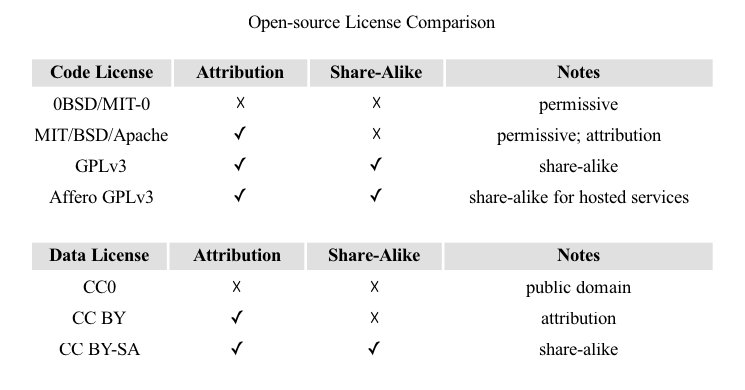}
\caption{Open licenses apply to software and data and can require attribution, or openness of derivative work. Permissive licenses assign copyright, but give users broad usage and distribution rights whereas works placed in the public domain release copyright from the original author. Creative Commons licenses are popular for data and content, but open-source code is more often licensed under one of the other listed licenses.}
\label{fig:licenses}
\end{figure}

When choosing a license, there are a few important considerations to make \cite{engelfriet_choosing_2010}. 
There may be requirements set by government or foundational funding sources requiring research outputs be released under specific terms \cite{wellcome_open_2025, gates_foundation_2025_nodate, hhmi_presidents_office_open_2020}.
Specific archives may also have license requirements—for example, Dryad places all depositions under a CC0 license.
After fulfilling any obligations, the intended audience of the code and data should be considered. 
Results that are intended to be built upon in a research setting may be best served under a share-alike license to encourage future openness. 
To maximize reach across commercial settings, placing code under permissive licenses and data in the public domain is recommended.
Conversely, a noncommercial license may be used if the goal is to discourage commercial use outside of paid license agreements.
Dual licensing may also be an option for authors that wish to retain the ability to license their work commercially under stricter terms \cite{valimaki_dual_2002}.
For research outputs, it is generally best to use a license that requires attribution at a minimum to preserve the historical record.
An overview of recommended open licenses is provided in Figure \ref{fig:licenses}.

\section{Discussion}

\subsection{Recommendations}

The following recommendations apply to specific groups of stakeholders involved with practicing, archiving, sharing, and funding research.

\subsubsection{Researchers}

Researchers should work towards sharing the computing environments that they perform their research in.
They should fully document their workflows which includes spelling out all dependency versions used, including packages and libraries used in code, programming languages, and operating systems tested on.
Shell scripts that define and run the workflow should also be provided.
Finally, providing a container image that bundles all aspects of the experiment is necessary in order to create a community around reproducing results and sharing scientific knowledge as broadly as possible. 
Implementing the ARTS open framework is a good way to achieve this and for well-designed computational studies, should primarily require specifying a \texttt{config.env} file and modifying associated top-level scripts.

\subsubsection{Archivists} 
 
Those who administer deposition archives should make sure that they have at least the minimal feature set described by Figure \ref{fig:archives}.
Except for hosting container registry endpoints, a few archives are close to being full-featured in this regard.
Zenodo and the Dataverse Project should expose file structure in URLs to support git repository cloning.
The Internet Archive could be a suitable archive for research projects if it added support for immutable releases and temporary private uploads for the purpose of peer review.
Stanford's Digital Repository would also support almost all aspects of the standard if it opened up submissions to external collaborators.
Incorporating useful provenance tools like trusted timestamping also moves some of the work off of researchers and increases the reputability of the archive.
All major archiving platforms should support pulls for OCI container images.

\subsubsection{Publishers}

 Publishers and those involved with the scientific review process should require that the research they are distributing conforms to reproducibility best practices.
 The peer review process for submissions that include code should involve having an independent recreation of relevant figures from referenced data.
 A number of Nature Publishing Group journals have taken a step in this direction by integrating and encouraging the use of Code Ocean for reviews \cite{nature_methods_easing_2018, nature_computational_science_seamless_2022}.
Publishers, especially those already partnered with archive services, should also be aware of the opportunity to implement automated ``clearing house'' services that run the deposited container to validate figure generation from data for ARTS-compatible submissions.
Encouraging third-party figure recreation in this manner sets a high bar for computational reproducibility.

\subsubsection{Funding agencies and policy makers}

Many US funding agencies are taking steps to ensure that data generated from sponsored awards be made public at the time of publication \cite{nelson_ensuring_nodate}.
While this is an excellent first step, simply having the data released does not mean the data is usable or accessible.
Ideally, code and the computational environment needed to generate the outputs of research should also be required to be made public in the same manner, and this would significantly increase the accessibility of research works.
The ARTS framework would be an example framework compliant with both the letter and the spirit of such requirements.

\subsection{Limitations and Future Work}

\subsubsection{Limitations}

The ARTS open framework is not a silver bullet for research reproducibility and replicability, but simply a set of flexible best practices. 
It is not a replacement for careful data collection, statistical rigor, or clear explanation of methods. 
The framework is dependent on a number of other technologies, notably: git (although it may be extensible to other version control systems), the OCI's container specifications, and chosen archive services. 
It also does not yet adequately address workflows that work with large raw data on the order of tens of terabytes or more per experiment or those that involve specific or proprietary hardware.

And although the ARTS open framework offers wide support across compute environments, steps to perform GPU workflows or hybrid workflows that span local and cloud compute clusters are not prescribed in this writing or accompanying code. 
Further, container images generated on one host architecture may not always reliably run on other host architectures \cite{scott_mccarty_limits_nodate}.

It should also be noted that container images for scientific computing tend to be quite large (potentially up to tens of gigabytes depending on the included dependencies).
Archives serving these files can take advantage of layer caching—the process of sharing reusable blocks of containers that share dependencies—to greatly reduce the total storage needed to host container images.

\subsubsection{Future Work}

ARTS is a generic framework, but to make adoption easier and more accessible for researchers in their respective fields, it's important to have opinionated examples.
One signal processing example experiment using the framework is provided at the time of writing, but for the ARTS open framework to be most useful, there must be further examples created that are compatible with the framework that detail specific data collection pipelines and shared environments for analyses using programming languages besides Python, such as R and Julia.
Examples of and documentation for ARTS-compatible depositions and the associated upload process across different archives including depositing within a cloud storage bucket would also make the framework easier to use.
Clear examples on how to archive large data on the order of terabytes with trusted timestamping and version control need also be added.

To improve adoption of archiving computation environments among publishers, an automated figure validation framework is needed which will run the \verb|ARTS_RUN_CMD| using the provided container image.
This service should be exposed initially as an open source infrastructure as code (IaC) template to encourage standardization across publishers.

\section{Methods}

\subsection{Example Implementation: Watch Calibration}

As a simple example experiment and computational analysis to demonstrate how the ARTS open framework is implemented, audio from a Chinese Standard Movement \cite{chinese_watch_wiki_chinese_nodate, wikipedia_chinese_2024} watch was recorded with a laptop microphone in order to determine the clock drift and calibrate the watch. 
This experiment was chosen since it highlights all aspects of the framework while offering an accessible way to replicate the experiment by generating a new dataset to anyone with a computer microphone and mechanical watch.

A 2813 automatic mechanical movement was recorded with a laptop by placing the movement and case on the interior of the laptop above the headphone jack where the microphone is located. 
Other watch movements and recording equipment may be used, but when recording with a laptop the best results are obtained with the watch directly on top of the laptop microphone and a quiet recording environment. 
Audio data was recorded at 44100 Hz using the sounddevice Python package. 
The experiment code also offers the ability to generate a dataset using the \texttt{librosa.clicks} method \cite{mcfee_librosalibrosa_2024}.
Derivative data in the form of audio windows containing each clock tick and their respective peak onset time were output to the \verb|ARTS_DERIV_DATA_PATH| folder. 

\begin{figure}[htb!]
\centering
\includegraphics[scale=0.3]{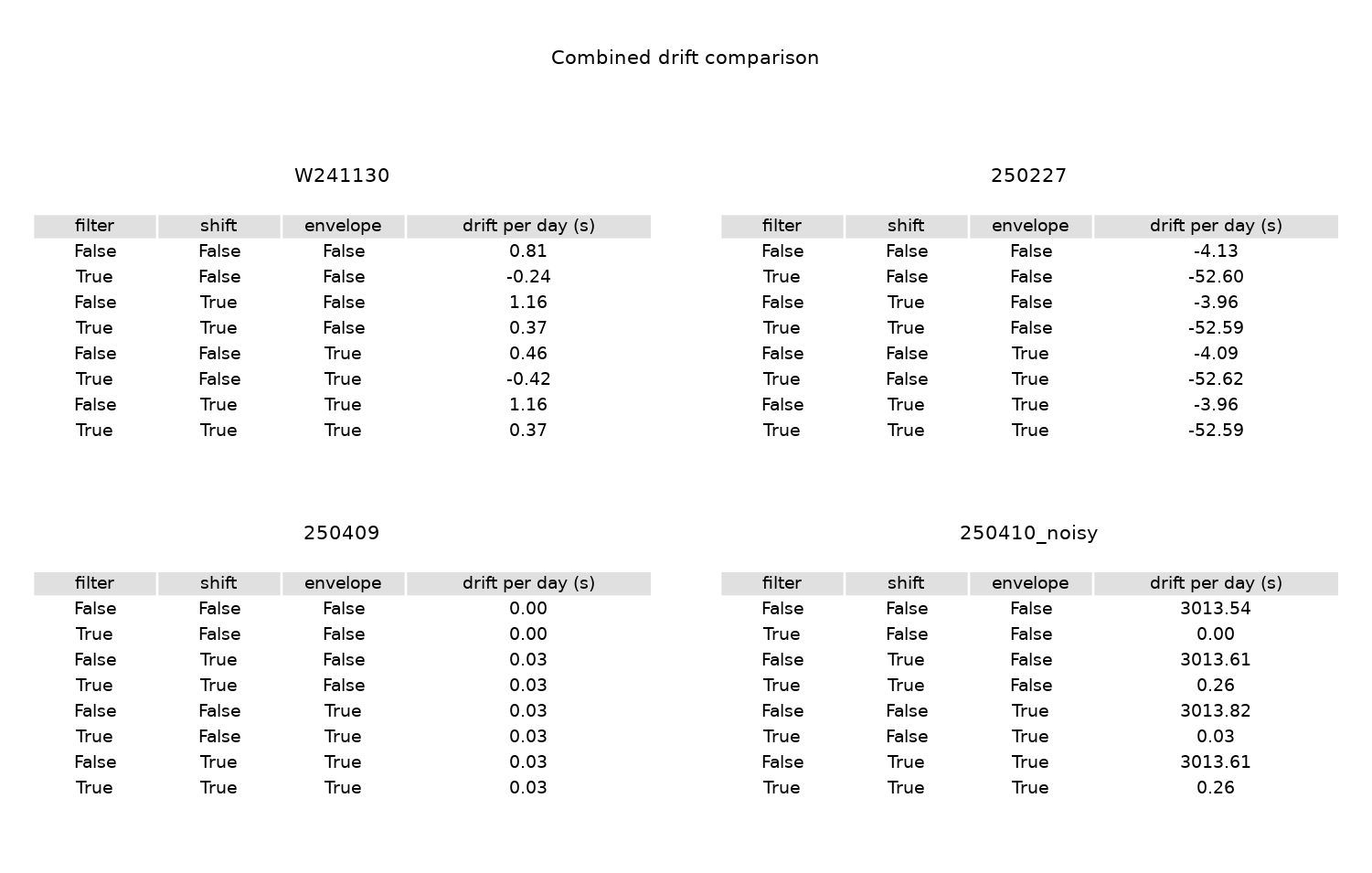}
\caption{Comparison of drift values across each dataset recorded (W241130; 250227) and generated (250409; 250310\_noisy) for all possible analysis parameters. Drifts are fairly consistent for recordings on the left with high SNR. Data with more noise may need filtering for a reasonable drift calculation (250410\_noisy) or filtering may result in additional noise (250227).}
\ref{fig:drift_figure}
\label{fig:drift_table}
\end{figure}

To calculate the drift of the watch movement over time, the audio signal peaks were picked out using \texttt{scipy.signal.find\_peaks}. 
Clock tick windows were then defined and adjacent windows were correlated with one another to adjust for noise in calculating the local maxima. 
Prominence parameters for \texttt{find\_peaks} were tuned to work using an audio signal with a high signal-to-noise ration, so noisy audio data may require additional filtering or optional usage of an envelope calculation included as parameters in the analysis.

\begin{figure}[!htb]
\centering
\includegraphics[scale=0.6]{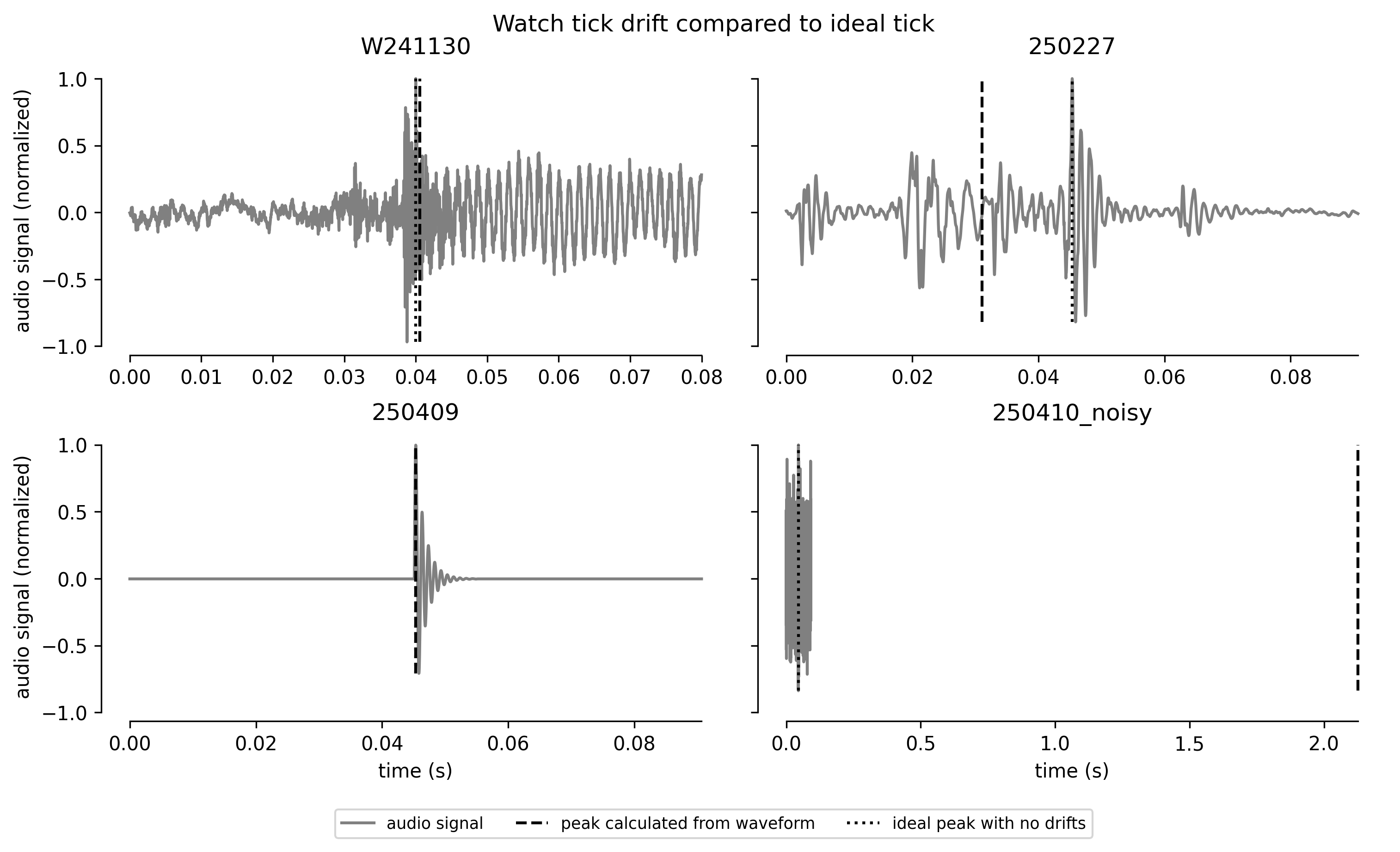}
\caption{Visualization of difference between a tick at the end of the recording (dashed line) and a corresponding tick from an ideal generated source (dotted line) as in the case of no filtering. The difference between these tick times corresponds to the first row in Figure 
\ref{fig:drift_table}}
\label{fig:drift_figure}
\end{figure}

The analysis can be run in its container environment via Podman or Docker using the \textbf{./run.sh} script available in the deposition \cite{dasgupta_watch_2025}. 
The run script provides a command to regenerate figures (\textbf{./run.sh generate-figures}), launch a jupyter notebook (\textbf{./run.sh jupyter}), or launch an ipython interactive shell (\textbf{./run.sh ipython}). 
The recording process can also be rerun using the \textbf{collect\_data.sh} script to collect new data using a system microphone or generate new data using librosa.
The container image may also be regenerated with \textbf{./run.sh save-image}.

\section{Data Availability}

All raw data used in the example experiment is available in the SDR (https://doi.org/10.25740/vs897sz1847, https://purl.stanford.edu/vs897sz1847), on Zenodo (https://zenodo.org/records/15232081), or in the Internet Archive (https://purl.archive.org/watch-calibration).

\section{Code Availability}

All code used in the example experiment is available in the SDR (https://doi.org/10.25740/vs897sz1847, https://purl.stanford.edu/vs897sz1847), on Zenodo (https://zenodo.org/records/15232081), or on GitHub (https://github.com/bil/watch-calibration).

The full git repository can be cloned with \texttt{git clone \url{https://stacks.stanford.edu/file/vs897sz1847/watch-calibration/watch-calibration.git}}, but data must be downloaded separately. Note that this link is not a PURL, but the root URL may be found by navigating to the SDR deposition DOI/PURL and copying the download link for an individual file.

\section{References}

\Urlmuskip=0mu plus 1mu\relax
\bibliographystyle{references}
\bibliography{references}

\section{Author Contributions}


Contributions are listed in accordance with the CRediT taxononmy. 
S.D.: Conceptualization, Data curation of all datasets except W241130, Formal Analysis, Investigation, Methodology, Software, Validation, Visualization, Writing—original draft, Writing—review and editing.
P.N.: Conceptualization, Data Curation of W241130 trusted timestamping dataset \cite{nuyujukian_w241130_2024}, Funding Acquisition, Methodology, Resources (GCP), Supervision, Writing—review and editing.

\section{Competing Interests}

The authors declare no competing interests.

\section{Acknowledgments}

The authors would like to thank the Brain Interfacing Lab group members for helpful feedback, Bryce Grier for thoughtful comments, the Wu Tsai Neurosciences Institute for support, and the NIH for funding via grant NIH U19NS118284.

\end{document}